\documentclass[showpacs,preprintnumbers,prl,twocolumn]{revtex4}%
\usepackage{amssymb}
\usepackage{amsmath}
\usepackage{graphicx}
\usepackage{bm}
\usepackage{amsfonts}%
\begin{document}
%

\title
[Short title for running header]{Decoherence in Josephson Qubits from Junction Resonances}%
%

\author
{R. W. Simmonds, K. M. Lang, D. A. Hite, D. P. Pappas, and John M.  Martinis}%
%

\email{martinis@boulder.nist.gov}%
%

\affiliation
{National Institute of Standards and Technology, 325 Broadway, Boulder, CO 80305-3328, USA}%
%

%

%

\begin{abstract}
Although Josephson junction qubits show great promise for quantum computing,
the origin of dominant decoherence mechanisms remains unknown. \ We report
Rabi oscillations for an improved phase qubit, and show that their
\textquotedblleft coherence amplitude\textquotedblright\ is significantly
degraded by spurious microwave resonators. \ These resonators arise from
changes in the junction critical current produced by two-level states in the
tunnel barrier. The discovery of these high frequency resonators impacts the
future of all Josephson qubits as well as existing Josephson technologies.
\ We predict that removing or reducing these resonators through materials
research will improve the coherence of all Josephson qubits. \
\end{abstract}%
%

\volumeyear{year}%
%

\volumenumber{number}%
%

\issuenumber{number}%
%

\eid{identifier}%
%

\date{\today}%
%

\maketitle

\ Josephson junctions are good candidates for quantum
computing\cite{QuantumComputing}, with recent experiments demonstrating
reasonably long coherence times, state preparation, manipulation, and
measurement, and the coupling of qubits for eventual gate
operations\cite{Vion02,Yu02,Martinis02,Chiorescu03,NEC03,Maryland03}.
\ Josephson quantum bits (qubits) may be considered as non-linear
\textquotedblleft LC\textquotedblright\ resonators formed by the Josephson
inductance and capacitance of a tunnel junction\cite{Martinis02}. \ Making
qubits from such electrical elements is advantageous because coupling of
qubits and scaling to large numbers should be relatively straightforward using
integrated-circuit fabrication technology. \ Although circuits have been
invented to decouple the qubit from unwanted electromagnetic modes,
solid-state systems are inherently complex and may contain defects that
degrade the coherence. \ A full knowledge of the system's Hamiltonian requires
probing the operational space of the qubit and investigating all decoherence
behavior. \ \

Here we report the discovery of spurious microwave resonators that reside
within the Josephson tunnel barrier used to form a solid-state qubit. \ We
show that at certain bias points, strong coupling to these resonators destroys
the coherence of the qubit. \ Although previous reports have focused on the
characteristic decay \textit{time} of coherent oscillations, our data
demonstrates that decoherence from these spurious resonators primarily affects
the \textit{amplitude} of the oscillations. \ We present a model that explains
these resonators as being produced by fluctuations in the tunnel barrier.
\ This phenomena can be compared with previous measurements of both junction
current-voltage characteristics and low frequency $1/f$ critical-current
noise. \ The sensitivity of our Josephson qubit at the quantum level has
allowed us to uncover these two level systems hidden within a forty year-old
technology. \ As major sources of decoherence, they play a vital role in
understanding the full effective Hamiltonian of Josephson tunnel junctions. \ \ \

The circuit used in this experiment, shown in Fig.\thinspace1(a), is cooled to
20 mK. \ The junction is isolated from dissipation of the leads in a similar
manner to a phase qubit described previously\cite{Martinis02}. \ The circuit
has been improved by placing the junction in a superconducting loop of
inductance $L$ to minimize the voltage and thus the generation of
quasiparticles and self-heating when the qubit state is measured\cite{Lang03}.
\ The junction is biased with current $I$ close to the critical current
$I_{0}$ by coupling magnetic flux through a transformer with mutual inductance
$M$. \ As shown in Fig.\thinspace1(b), the qubit states are formed in a cubic
potential of the left well and are measured by tunneling to states in the
right well. \ Tunneling to the right well changes the flux through the loop by
$\sim\Phi_{0}=h/2e$, which is easily read-out with a pulsed critical-current
measurement in a separate SQUID detector. \ The qubit was fabricated using
aluminum metallization, with an aluminum-oxide tunnel barrier formed by an
ion-mill clean followed by thermal oxidation\cite{Lang03}.%

\begin{figure}
[pb]
\begin{center}
\includegraphics[
height=1.5895in, width=3.3684in
]%
{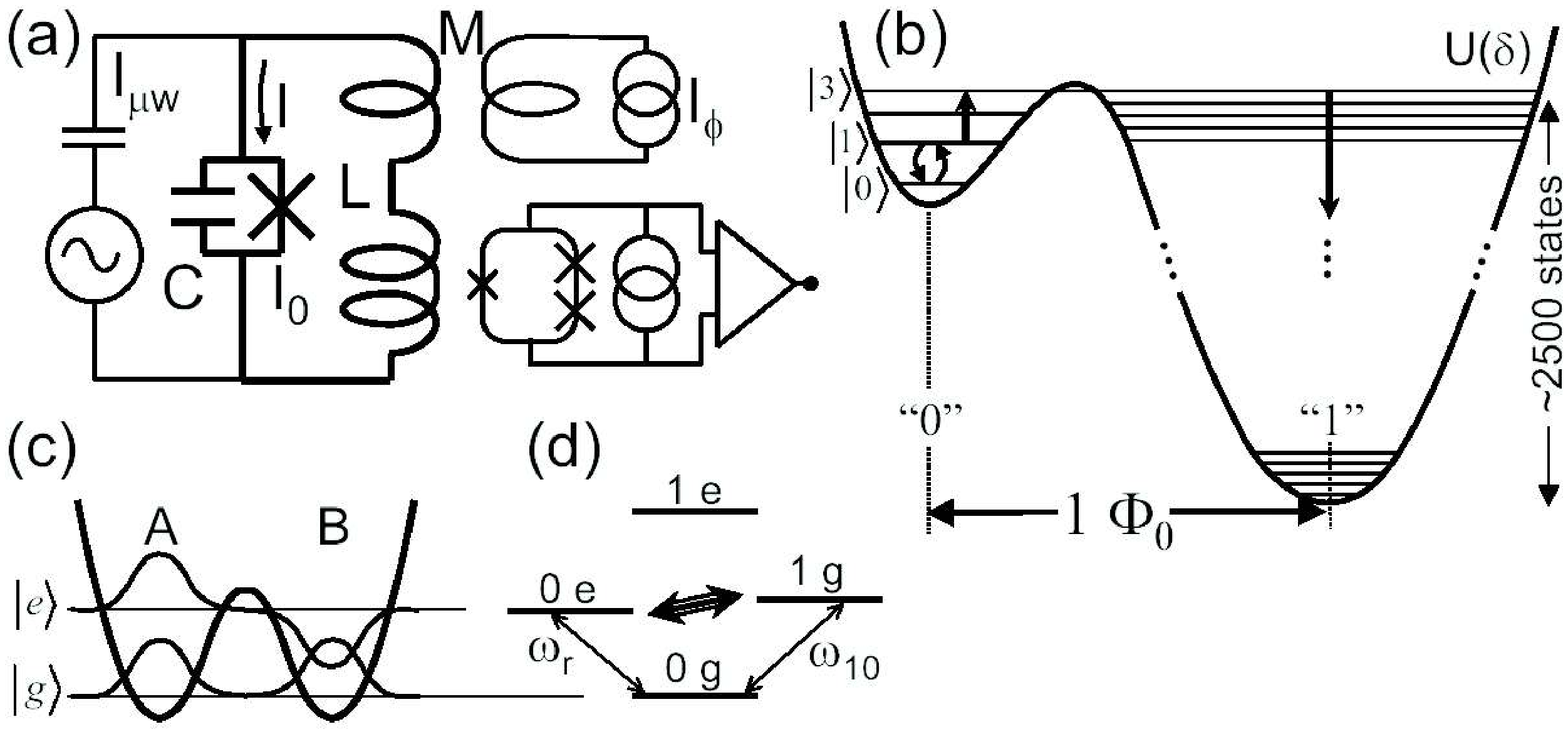}%
\caption{(a) Circuit diagram for the Josephson junction qubit. \ Junction
current bias $I$ is set by $I_{\phi}$ and microwave source $I_{\mu\mathrm{w}}%
$. \ Parameters are $I_{0}\simeq11.659\,\mu\mathrm{A}$, $C\simeq
1.2\,\mathrm{pF}$, $L\simeq168\,\mathrm{pH}$, and $L/M\simeq81$. \ (b)
Potential energy diagram of qubit, showing qubit states $\left\vert
0\right\rangle $ and $\left\vert 1\right\rangle $ in cubic well at left.
\ Measurement of $\left\vert 1\right\rangle $ state performed by driving the
$1\rightarrow3$ transition, tunneling to right well, then relaxation of state
to bottom of right well. \ Post-measurement classical states \textquotedblleft%
0\textquotedblright\ and \textquotedblleft1\textquotedblright\ differ in flux
by $\Phi_{0}$, which is readily measured by readout SQUID. \ (c) Schematic
description of tunnel-barrier states A and B in a symmetric well. \ Tunneling
between states produces ground $\left\vert g\right\rangle $ and excited
$\left\vert e\right\rangle $ states separated in energy by $\hbar\omega_{r}$.
\ (d) Energy-level diagram for coupled qubit and resonant states for
$\omega_{10}\simeq\omega_{r}$. \ Coupling strength between states $\left\vert
1g\right\rangle $ and $\left\vert 0e\right\rangle $ is given by $\widetilde
{H}_{int}$. }%
\end{center}
\end{figure}

The $0\rightarrow1$ qubit transition frequency $\omega_{10}$ is measured
spectroscopically\cite{elevels}, as shown in Fig.\thinspace2(a). \ For our
measurements, the current bias is pulsed for a time $\sim50\,\mu\mathrm{s}$ to
a value close enough to the critical current so that approximately $3-4$
energy levels are in the cubic well. \ Transition frequencies are probed by
applying microwave current $I_{\mu\mathrm{w}}$ at frequency $\omega$ and
measuring a resonant increase in the net tunneling probability. \ We observe
in Fig.\thinspace2(a) a decrease in the transition frequency as the bias
current approaches the critical current, as expected theoretically for the
$\omega_{10}$ transition. \

Additionally, we observe a number of small spurious resonances (indicated by
dotted vertical lines) that are characteristic of energy-level repulsion
predicted for coupled two-state systems. \ These extra resonances have a
distribution in splitting size, with the largest one giving a splitting of
$\sim25\,\mathrm{MHz}$ and an approximate density of $1$ major spurious
resonance per $\sim60\,\mathrm{MHz}$. \ \ %

\begin{figure}
[pb]
\begin{center}
\includegraphics[
height=5.1863in,
width=3.3313in
]%
{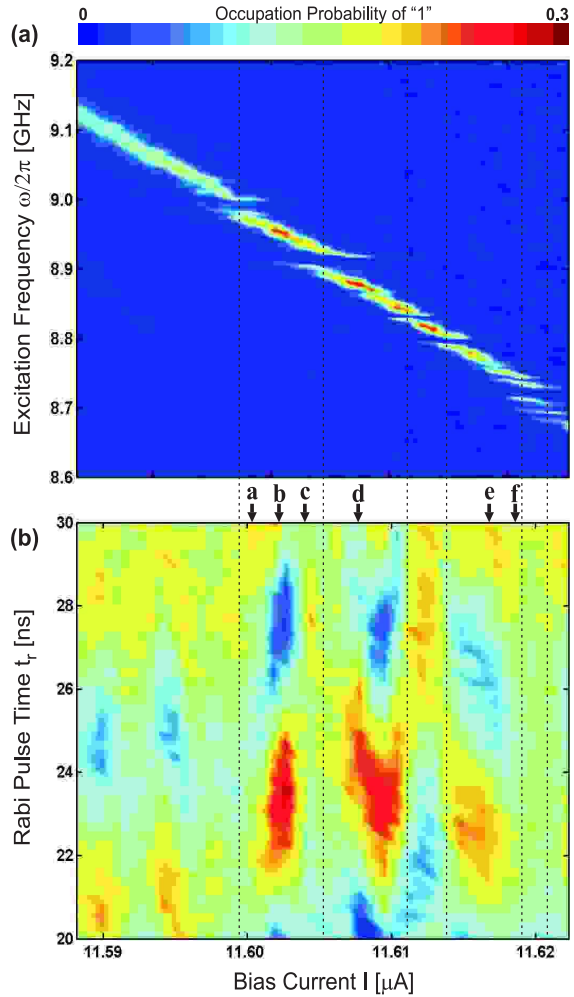}%
\caption{(a) Measured probability of state \textquotedblleft%
1\textquotedblright\ versus microwave excitation frequency $\omega/2\pi$ and
bias current $I$ for a fixed microwave power. \ Data indicate $\omega_{10}$
transition frequency. \ \ Dotted vertical lines are centered at spurious
resonances. \ (b) Measured occupation probability of $\left\vert
1\right\rangle $ versus Rabi-pulse time $t_{r}$ and bias current $I$. \ In
panel (b), a color change from dark blue to red corresponds to a probability
change of 0.4. \ Color modulation in time $t_{r}$ (vertical direction)
indicates Rabi oscillations. \ }%
\end{center}
\end{figure}

We observed coherent \textquotedblleft Rabi oscillations\textquotedblright%
\ between the $\left\vert 0\right\rangle $ and $\left\vert 1\right\rangle $
state by pulsing microwaves at the $0\rightarrow1$ transition frequency
$\omega_{10}$, then measuring the occupation probability of state $\left\vert
1\right\rangle $ by applying a second microwave pulse resonant with the
$1\rightarrow3$ transition frequency $\omega_{31}$\cite{Martinis02}. \ Figure
3(a)-(c) shows for three values of microwave power the occupation probability
versus the Rabi pulse time $t_{r}$ at a bias point away from any resonances.
\ The decay of the oscillations is approximately exponential and gives a
coherence time of $41\,\mathrm{ns}$. \ Figure\thinspace3(d) shows that the
oscillation frequency is proportional to the microwave amplitude, as expected. \ \ %

\begin{figure}
[pb]
\begin{center}
\includegraphics[
height=2.2157in,
width=3.0796in
]%
{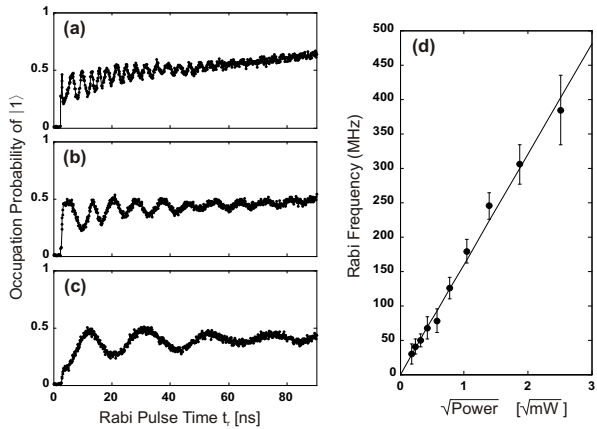}%
\caption{(a)-(c) Measured occupation probability of $\left\vert 1\right\rangle
$ versus time duration of Rabi pulse $t_{r}$ for three values of microwave
power, taken at bias $I=11.609\,\mu\mathrm{A}$ in Fig.\thinspace2. \ The
applied microwave power for (a), (b), and (c) correspond to $0.1$, $0.33$, and
$1.1\,\mathrm{mW}$, respectively. \ (d) Plot of Rabi oscillation frequency
versus microwave amplitude. \ A linear dependence is observed, as expected
from theory. \ }%
\end{center}
\end{figure}

The correlation between Rabi oscillations and spurious resonances is
demonstrated in Fig.\thinspace2. \ For the data in Fig.\thinspace2(b), the
microwave frequencies of the Rabi and measurement pulses were adjusted with
$I$ to center on the transition frequencies obtained from spectroscopy data.
\ We observe that the oscillation amplitude, represented by a variation in
color, is suppressed over time (vertical axis) at particular bias currents
(horizontal axis). \ The dashed vertical lines show that this suppression is
correlated to those transition frequencies $\omega_{10}$ where there are
pronounced resonance structures. \ It is clear that these spurious resonances
strongly disrupt the Rabi oscillations. \ \

In Fig.\thinspace4 we show the decay of the Rabi oscillations for biases
indicated by arrows in Fig.\thinspace2(b). \ Near resonances we find unusual
behavior such as beating (b), loss and recovery of the oscillations with time
(c), and rapid loss of coherence amplitude (f). \ The general trend is that
spurious resonances cause loss in coherence not by a decrease in the decay
time of the oscillations, but by a decrease in the amplitude of the
oscillations. \ %

\begin{figure}
[pb]
\begin{center}
\includegraphics[
height=2.4915in,
width=3.0813in
]%
{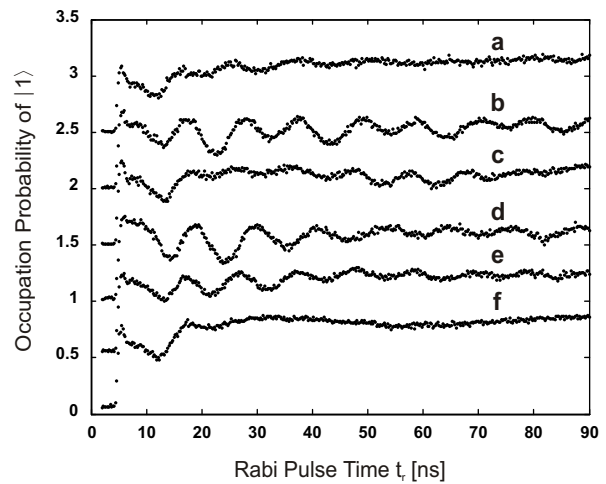}%
\caption{Measured occupation probability of $\left\vert 1\right\rangle $
versus time duration of Rabi pulse $t_{r}$ for current biases a-f as noted by
arrows in Fig. 2. \ Data a-e is offset for clarity. \ Note that when the bias
is changed, the coherence is degraded mainly as a loss in amplitude, not by a
decrease in coherence time. \ }%
\end{center}
\end{figure}

Experimental checks were performed to rule out whether these resonances arise
from coupling to the read-out SQUID or modes in the device mount\cite{checks}.
\ Additionally, simulations show that the beating behavior in Fig.\thinspace4
is consistent with the interaction of the qubit with another two level system
but not harmonic oscillator modes from the leads of the device. \ When
thermally cycled to room temperature, the magnitude and frequency of the
spurious resonances can change considerably, whereas cycling to 4K produces no
apparent effect. \ Furthermore, for the same experimental setup and over 10
qubit devices, we find that each qubit has its own unique ``fingerprint" of
resonance frequencies and splitting strengths. \ Finally, over many hours
while the device is cold, we observe that resonance frequencies and strengths
can change spontaneously. \ All of these observations indicate that the
resonances are microscopic in origin. \ \

We have constructed a model for the microscopic resonators, where we consider
two-level states in the barrier with large tunneling matrix elements
corresponding to a microwave frequency\cite{trapnote}. \ If we consider two
states in the tunnel barrier that have configurations $A$ and $B$ producing
critical currents $I_{0A\text{ }}$and $I_{0B}$, then the interaction
Hamiltonian between the resonance and the critical current is%
\begin{align*}
H_{int}=  &  -\frac{I_{0A}\Phi_{0}}{2\pi}\cos\widehat{\delta}\otimes\left\vert
\Psi_{A}\right\rangle \left\langle \Psi_{A}\right\vert \\
&  -\frac{I_{0B}\Phi_{0}}{2\pi}\cos\widehat{\delta}\otimes\left\vert \Psi
_{B}\right\rangle \left\langle \Psi_{B}\right\vert \text{ ,}%
\end{align*}
where $\widehat{\delta}$ is an operator corresponding to the phase difference
across the junction, and $\Psi_{A,B}$ describe the two wave functions for the
two configuration states within the tunnel barrier. \ If we assume a symmetric
potential with energy eigenstates separated by $\hbar\omega_{r}$, the ground
and excited states are given by $\left\vert g\right\rangle \simeq(\left\vert
\Psi_{A}\right\rangle +\left\vert \Psi_{B}\right\rangle )/\sqrt{2}$ and
$\left\vert e\right\rangle \simeq(\left\vert \Psi_{A}\right\rangle -\left\vert
\Psi_{B}\right\rangle )/\sqrt{2}$. \ Using matrix elements for $\cos
\widehat{\delta}$ appropriate for the phase qubit\cite{Martinis03a} and
including only the dominant resonant terms arising from this interaction
Hamiltonian, we find%
\[
\widetilde{H}_{int}=\frac{\Delta I_{0}}{2}\sqrt{\frac{\hbar}{2\omega_{10}C}%
}\left(  \left\vert 0\right\rangle \left\langle 1\right\vert \otimes\left\vert
e\right\rangle \left\langle g\right\vert +\left\vert 1\right\rangle
\left\langle 0\right\vert \otimes\left\vert g\right\rangle \left\langle
e\right\vert \right)  \text{ ,}%
\]
where $\Delta I_{0}=I_{0A}-I_{0B}$. \ Figure 1(d) shows an energy level
diagram for the case where $\omega_{r}\approx\omega_{10}$. \ The coupling of
the two intermediate energy levels through $\widetilde{H}_{int}$ produces a
repulsion in the energy eigenstates that corresponds to the spectroscopic data
in Fig.\thinspace2(a). \ From the magnitude of level repulsions at resonance,
the largest value being $2|\widetilde{H}_{int}|/h\simeq25\,\mathrm{MHz}$, we
obtain from our model, $(\Delta I_{0}/I_{0})_{res}\lesssim65\cdot10^{-6}$. \

If we assume that the two configurations $A$ and $B$ correspond to turning on
and off an individual conduction channel in the tunnel barrier, then $(\Delta
I_{0}/I_{0})$ can be estimated from the current-voltage characteristics of the
junction. \ For a non-uniform tunnel barrier, mesoscopic
theory\cite{MARtheory} predicts that the total current flowing through a
junction comes from a sum over a number of conduction channels with tunneling
transmissions $\tau_{i}$. These transmission coefficients can be measured from
steps in the subgap current-voltage characteristics\cite{Goffman00}. \ These
steps have magnitude $2/\tau_{i}$\ and arise from $n^{\mathrm{th}}$ order
multiple Andreev reflections at subharmonic gap voltages $2\Delta/en$. \ If we
assume the current is carried by $N_{ch}$ channels, each with an average
transmission $\tau$, then our measured current-voltage
characteristics\cite{Lang03} imply $\tau\approx4\cdot10^{-3}$ for a
critical-current density of $\sim40\,\mathrm{A}/\mathrm{cm}^{2}$. \ For our
qubit junction with a normal-state resistance $R_{N}=29\,\Omega$, we calculate
that the average fraction of the critical current carried by each channel is
roughly $(\Delta I_{0}/I_{0})_{I-V}=1/N_{ch}=2\tau R_{N}e^{2}/h\simeq
8\cdot10^{-6}$. \ This value is in reasonable agreement with those obtained
from the magnitudes of the resonance splittings considering that the junction
has a distribution of tunneling transmissions $\tau_{i}$ and our measurements
of the level repulsions only identify a small fraction of the largest
resonators from their full distribution.

We can also compare these results with $1/f$ critical-current noise
measurements at audio frequencies for individual $1/f$ fluctuators within
submicrometer junctions\cite{Rogers84,VanHarlingen02,wakai86}. \ For an
aluminum junction with area $0.08\,\mu\mathrm{m}^{2}$, a single
fluctuator\cite{VanHarlingen02} gave a change in critical current of $\Delta
I_{0}\simeq10^{-4}I_{0}$. \ Scaling this fractional change to our qubit
junction with an area of $32\,\mu\mathrm{m}^{2}$, we obtain $(\Delta
I_{0}/I_{0})_{1/f}\simeq0.3\cdot10^{-6}$ for a single fluctuator, about 30
times smaller than estimated above from the current-voltage characteristics.
\ A factor of $\sim4$ in this ratio is due to enhanced subgap current arising
from surface inhomogenieties created by an ion-mill cleaning step in our
junction fabrication process. \ Additionally, the density of fluctuators in
frequency may also be compared. \ Several experiments give approximately one
resonance per decade in frequency for junctions with area $0.1\,\mu
\mathrm{m}^{2}$\cite{Rogers84,VanHarlingen02}, which is higher by only a
factor of $2$ than the areal density we observed for the resonances at
microwave frequencies. \

The magnitude and density in frequency of the microwave resonances and low
frequency $1/f$ noise fluctuators are in reasonable agreement, especially
since they compare phenomena that have characteristic frequencies separated by
many orders of magnitude. \ With good agreement between the resonator model,
our data, and data from $1/f$ noise, it seems plausible that junctions with
low $1/f$ noise will contain weaker spurious microwave resonators. \ This
suggests that we can estimate from previous $1/f$ measurements the performance
of qubits using alternative materials. \ A recent compilation of $1/f$ noise
data indicates that tunnel junctions made from oxides of Al, Nb, and PbIn all
have similar magnitudes of noise\cite{VanHarlingen02}. \ This evidence
suggests that alternatives to simple thermal or plasma oxidation of metals
should be investigated. \

In conclusion, we have developed an improved Josephson-phase qubit whose
sensitivity at the quantum level has allowed the discovery of spurious
microwave resonators within Josephson tunnel junctions. \ We have observed a
strong correlation between these resonators and decoherence in the qubit,
showing that they primarily degrade the coherence of Rabi oscillations through
a reduction of the \textquotedblleft coherence amplitude.\textquotedblright%
\ \ These resonators can be modeled as arising from two-level states within
the tunnel barrier, which couple to the qubit's states through the critical
current. \ This represents a new term in the Hamiltonian of fabricated
Josephson tunnel junctions that has remained hidden for over 40 years. \ These
microwave resonators may be affecting the performance of existing Josephson
technologies. Furthermore, we predict that improvements in the coherence of
all Josephson qubits will require materials research directed at
redistributing, reducing, or removing these resonance states. \ \ \

We thank M. Devoret, R. Koch, and V. Shumeiko for helpful discussions. \ This
work is supported in part by NSA\ under contract MOD709001.


\begin{thebibliography}{99}
\bibitem {checks}The frequencies of resonant modes in our mounting box is
consistent with its dimensions, and is not correlated with the spurious
resonances. \ Additionally, the resonance frequency of the measurement SQUID
depends on its bias current - no change in the spurious frequencies was
observed as this bias was varied. \
\bibitem {trapnote}Critical-current fluctuations are often depicted as arising
from trapping of electrons in the barrier (see for example Ref.
\cite{VanHarlingen02}). \ With experimental data showing non-uniform barriers
and no qualitative change in the $1/f$ noise below $T_{c}$, we believe our
tunneling-channel model is more physically realistic. \
\end{thebibliography}
\end{document}